\documentstyle[12pt,epsf,epsfig]{article}
\newfont{\thiplo}{msbm10 scaled\magstep 2}
\newfont{\gothic}{eufb10 scaled\magstep 2}
\newfont{\unc}{eurb10} 
\newskip\humongous \humongous=0pt plus 1000pt minus 1000pt
\def\caja{\mathsurround=0pt}\def\eqalign#1{\,\vcenter{\openup1\jot \caja
        \ialign{\strut \hfil$\displaystyle{##}$&$ 
        \displaystyle{{}##}$\hfil\crcr#1\crcr}}\,}
\newif\ifdtup

\def\eqright #1\cr{\noalign{\hfill$\displaystyle{{}#1}$}}
\def\eqleft #1\cr{\noalign{\noindent$\displaystyle{{}#1}$\hfill}}

\def\oldreffmt#1{\rlap{[#1]} \hbox to 2\parindent{}}

\def\figfmt#1{\rlap{Figure {#1}} \hbox to 1in{}}

\def\sectioneq{\def\theequation{\thesection.\arabic{equation}}{\let
\holdsection=\section\def\section{\setcounter{equation}{0}\holdsection}}}%

\newcounter{holdequation}

\def\begineq #1\endeq{$$ \refstepcounter{equation}\eqalign{#1}\eqno
	(\theequation) $$}
\def\contlimit{\,{\hbox{$\longrightarrow$}\kern-1.8em\lower1ex
\hbox{${\scriptstyle (a\rightarrow0)}$}}\,}
\def\centeron#1#2{{\setbox0=\hbox{#1}\setbox1=\hbox{#2}\ifdim
\wd1>\wd0\kern.5\wd1\kern-.5\wd0\fi
\copy0\kern-.5\wd0\kern-.5\wd1\copy1\ifdim\wd0>\wd1
\kern.5\wd0\kern-.5\wd1\fi}}
\def\centerover#1#2{\centeron{#1}{\setbox0=\hbox{#1}\setbox
1=\hbox{#2}\raise\ht0\hbox{\raise\dp1\hbox{\copy1}}}}
\def\centerunder#1#2{\centeron{#1}{\setbox0=\hbox{#1}\setbox
1=\hbox{#2}\lower\dp0\hbox{\lower\ht1\hbox{\copy1}}}}
\def\lsim{\;\centeron{\raise.35ex\hbox{$<$}}{\lower.65ex\hbox
{$\sim$}}\;}
\def\gsim{\;\centeron{\raise.35ex\hbox{$>$}}{\lower.65ex\hbox
{$\sim$}}\;}

\def\super#1{\ifmmode \hbox{\textsuper{#1}}\else\textsuper{#1}\fi}
\def\textsuper#1{\newcount\holdspacefactor\holdspacefactor=\spacefactor
$^{#1}$\spacefactor=\holdspacefactor}

\def\getcite#1,{\advance\citenumber by1
\def\getcitearg{#1}\def\lastarg{@}
\ifnum\citenumber=1
\ref{#1}\let\next=\getcite\else\ifx\getcitearg\lastarg\let\next=\relax
\else ,\ref{#1}\let\next=\getcite\fi\fi\next}

\def\pom{{\rm P\kern -0.53em\llap I\,}}
\def\spom{{\rm P\kern -0.36em\llap \small I\,}}
\def\sspom{{\rm P\kern -0.33em\llap \footnotesize I\,}}

\relax
\def\contlimit{\,{\hbox{$\longrightarrow$}\kern-1.8em\lower1ex
\hbox{${\scriptstyle (a\rightarrow0)}$}}\,}
\def\upon #1/#2 {{\textstyle{#1\over #2}}}
\relax

\sectioneq

\def\til#1{\centeron{\hbox{$#1$}}{\lower 2ex\hbox{$\char'176$}}}
\def\tild#1{\centeron{\hbox{$\,#1$}}{\lower 2.5ex\hbox{$\char'176$}}}
\def\sumtil{\centeron{\hbox{$\displaystyle\sum$}}{\lower
-1.5ex\hbox{$\widetilde{\phantom{xx}}$}}}

\begin{document} 

\begin{titlepage} 

\rightline{\vbox{\halign{&#\hfil\cr
&\today\cr}}} 
\vspace{0.25in} 

\begin{center} 

\centerline{The Gribov Legacy, Gauge Theories and the Physical S-Matrix}

\medskip

Alan R. White\footnote{arw@anl.gov }

\vskip 0.6cm

\centerline{Argonne National Laboratory}
\centerline{9700 South Cass, Il 60439, USA.}
\vspace{0.5cm}

\end{center}

\begin{abstract}Reggeon unitarity and non-abelian gauge field copies are focussed on as two Gribov discoveries that, it is suggested, may ultimately be seen as the most significant and that could, in the far distant future, form the cornerstones of his legacy. The crucial role played by the Gribov ambiguity in the construction of gauge theory bound-state amplitudes via reggeon unitarity is described. It is suggested that the existence of a physical, unitary, S-Matrix in a gauge theory is a major requirement that could even determine the theory. 
\end{abstract} 
 
 \vspace{1in}
 
\noindent Contributed to the Gribov-85 Memorial Volume "Exploring Quantum Field Theory"

\end{titlepage}

\section{Introduction}

Vlodya Gribov was a brilliant theoretical physicist, ranking amongst the greatest of the twentieth century. The list of his discoveries and achievements is very long and is generally well-known. In this paper\footnote{I was very sorry to miss the Gribov-85 conference, primarily because I could not obtain a visa. I am very happy to have this opportunity to express my deep admiration for Gribov's work.} I will focus on just two discoveries that, from my own work, I think are the most significant and that I think could, in the far distant future, form the cornerstones of his legacy. They are reggeon unitarity and the Gribov ambiguity due to non-abelian gauge field copies. These discoveries were made well over 10 years apart and superficially involve very different physics. Both may be even more profound than is generally perceived and I will link them by elaborating the crucial role played by the Gribov ambiguity in my construction of gauge theory bound-state amplitudes via reggeon unitarity. I will suggest that
the existence of a physical, unitary, S-Matrix in a gauge theory could be a highly non-trivial constraint that may even determine the theory.    

\section{Reggeon Unitarity}

This was the outcome of a particularly intense collaboration between two equally brilliant physicists, Gribov and Pomeranchuk, during a period of several years shortly before
Pomeranchuk's death. It was a stunning leap forward. They took the low-order, already elaborate, t-channel unitarity calculations of feynman diagrams by Mandelstam and built a beautiful multiparticle complex angular momentum formalism\cite{gpt} that led to the existence of a unitarity condition for all multiple regge pole states. I will describe what I think are the most dramatic implications, not all of which are generally appreciated, but first I will discuss some early developments that followed the initial discovery.

\subsection{Analyticity Problems}

The GPT analysis\cite{gpt} made fundamental technical assumptions
about the complex angular momentum and helicity continuations of multiparticle amplitudes.
Subsequently, these assumptions seemed to be invalidated when the analyticity properties of the simplest (relevant) feynman production amplitudes were investigated\cite{ans}. Unfortunately, perhaps, this convinced Gribov that because of the complicated Landau singularity structure of multiparticle momentum space amplitudes, the complex angular momentum continuation of multiparticle t-channel unitarity could not work as assumed by GPT. As a result, he turned to field theory based models and formulated the Reggeon Calculus\cite{rc}, which was effectively an explicit solution of reggeon unitarity. Regrettably, the apparent model dependence of the formulation allowed the theory community at large to believe that the physical relevance of the reggeon calculus, and the equivalent reggeon field theory\cite{rft}, was an open question.

\subsection{Simplified Asymptotic Singularity Structure}

Gribov remained convinced, nevertheless, that even though the complex angular momentum and helicity continuations assumed might not exist, the GPT analysis was valid near the regge cuts (angular momentum plane branch-points) associated with multiple regge pole exchanges. He was unable to justify this but from my own work, in particular my collaboration with Henry Stapp on asymptotic dispersion relations\cite{hps}, I had understood that multiparticle complex angular momentum theory should be based on the simplified asymptotic analytic structure of many particle amplitudes in multi-regge regions. The asymptotic dispersion relations break up amplitudes into component parts that each have distinct asymptotic cut structures (as determined by the Steinmann relations or, in essence, according to the time-ordering of interactions). The component amplitudes then have distinct complex angular momentum and helicity continuations\cite{arw} that can be used in a more elaborate version of the GPT analysis. Ultimately, this analysis showed that Gribov's faith in the  GPT discontinuity formulae was justified.  
   
\subsection{Implications of Reggeon Unitarity}

The reggeon unitarity discontinuity formulae have an analagous structure to the momentum space unitarity equations and, similarly, are only well-defined when the basic complex angular momentum singularities are regge poles with factorizing residues. Extensive t-channel analytic continuation of the unitarity equations can then be used to show\cite{sim} that regge poles and regge cuts are the only singularities remaining (on the physical sheet) near $t = 0$ and so they control the s-channel high-energy limit. Most importantly, the complex angular momentum and helicity formalism based on asymptotic dispersion relations shows\cite{arw} that the reggeon unitarity equations extend to all multi-regge regions of multiparticle amplitudes. Consequently, if the basic angular momentum plane singularities of physical amplitudes are regge poles then a complete asymptotic solution of multiparticle t-channel unitarity can be obtained. 
In particular, reggeon unitarity can be applied to the study of bound-state amplitudes via multi-regge theory.

Basic singularities that are not regge poles are analogous to massless particles in momentum space in that they do not give well-defined 
multiple singularity contributions. As a result they are unable to provide an angular momentum plane solution of multiparticle t-channel unitarity. I believe that, eventually, the asymptotic dispersion relation based multi-regge theory that I have formulated will attract  substantial general interest, and will undergo the exposure and further development that is probably necessary for reggeon unitarity to become part of the established basis of particle physics theory. If this happens, the implications will be dramatic and will include the following.  

\subsubsection{The Physical Pomeron}

Reggeon unitarity implies that the leading singularity in the vacuum channel must be a 
regge pole(s) together with associated regge cuts. If total cross-sections do not fall asymptotically, there must be a vacuum regge pole with unit intercept and indeed, experimentally, there is very strong evidence\cite{tot,arw1} that the pomeron is approximately a single regge pole. In this case, all the associated regge cuts will also have unit intercept and the only known solution of reggeon unitarity is the Critical Pomeron.  

\subsubsection{Gauge Theories}

In a general non-abelian gauge theory, and QCD in particular, the conventional assumption is that perturbative parton model calculations, at large transverse momentum, can be extended into the regge region. It being assumed that confinement is a small transverse momentum phenomenon that can be ignored at sufficiently large transverse momentum. In leading order this leads to a two (gluon) reggeon cut which, after iteration of interactions (via a scale-invariant kernel) produces the BFKL pomeron as the leading singularity. As is widely known, it is an angular momentum cut\cite{fkl} which has no possibility to satisfy reggeon unitarity. Consequently, if the physical S-Matrix satisfies reggeon unitarity, it can not contain the BFKL pomeron as a physical singularity.

\subsubsection{Quantized Gravity?}

If the graviton is a regge pole, because it has spin two the regge pole intercept also has to be two. Reggeon unitarity then requires a two graviton regge cut with intercept three. In addition, multiple graviton regge cuts must also exist. They have analagous intercepts that are increasing integers and that will produce corresponding power increases of the total cross-section. Since reggeon unitarity is an exact S-Matrix property, it does not allow regge cut contributions to be removed by summation, for example, into an eikonal amplitude. Unavoidably, if the graviton is a regge pole, the forward amplitude is not polynomially bounded. Consequently, gravity can not be quantized and described by an S-Matrix with conventional analyticity properties, that satisfies reggeon unitarity. Most likely, from my viewpoint, gravity is not quantized.

\subsection{Massive Gauge Theory Reggeons}

In a spontaneously-broken gauge theory, with a suitably chosen symmetry breaking scalar sector, all the elementary gauge bosons and fermions acquire masses, while also becoming regge poles (reggeizing) in low-order perturbation theory. Therefore, reggeon unitarity should be satisfied and, indeed, as higher-orders are constructed (at high-energy), the multi-reggeon diagrams appear\cite{bs} that are predicted. Using the broad applicability of reggeon unitarity discussed above, the structure of reggeon diagrams in general multiparticle multi-regge limits can then be predicted. In principle, the contributions of arbitrarily high order feynman diagrams are included. In the next Section, I will discuss the massless limit for reggeon amplitudes and the pattern of infra-red divergences that occur. Because the regge region includes low transverse momenta we could expect to discuss possible bound-states that are a consequence of the divergences. Moreover, because of the range of multi-regge limits that can be considered, bound-state scattering amplitudes should also be accessible. 
As I will discuss in the next Section, the Gribov ambiguity, implied by the existence of gauge field copies, plays a crucial role in the massless limit and in the formation of bound-state scattering amplitudes.

\section{The Gribov Ambiguity}

When Gribov first described the phenomenon of non-abelian gauge field copies to a western audience (at the 1977 EPS conference in Budapest), it caused much excitement. Most western physicists were, of course, unaware that Gribov had previously announced\cite{gr} his discovery at the Leningrad Winter School. At first the copies seemed only to require a straightforward modification\cite{gr1} of the functional space in a gauge theory functional integral. Moreover, as Gribov himself suggested\cite{gr1}, it seemed that this might produce the confinement of color, via the growth of the coupling, that the theory community was desperate to discover. 

Unfortunately, perhaps, in the nearly forty years since Gribov's discovery, no satisfactory explicit proposal for how the functional integral is to be handled has appeared. Moreover, there is no acceptance of the idea that eliminating the copies will actually produce the conventionally desired confinement. Instead, 
the copies may actually imply that there is no well-defined non-perturbative definition of a functional integral giving correlation functions for a non-abelian gauge theory. In reality, as is well-known, the infinite volume divergence of the integral is already a big threat to it's definitive existence. However, without a well-defined function space, it seems that it can not provide even a formal vehicle for the non-perturbative definition of the theory. 

I will, instead, focus on Gribov's simple statement\cite{gr} of the basic problem - 

{\it ``in distinction to electrodynamics,
in non-Abelian theories it is not possible to uniquely introduce three-dimensional
transverse fields (in particular, transverse fields can be pure gauge).''} 

\noindent The vectors in massive reggeon diagrams contain three-dimensional transverse components and so if we can take the massless limit in a well-defined way we might obtain a unique result that satisfies reggeon unitarity. There are a variety of non-trivial issues involved in the taking of this limit and the crucial role of the Gribov ambiguity emerged (in my work) as a necessary component of a gauge theory origin for the Critical Pomeron. However, I will initially give a more general description of the phenomena involved.

\subsection{Partial Gauge Symmetry and Reggeon Anomaly Vertices}
 
With a transverse momentum cut-off, the general expectation would be that the reggeon massless limit will produce a cancellation of infra-red divergences in color zero channels and exponentiation (reggeization) of divergences in color non-zero channels. However, if massless fermions are present and we first {\bf restore only} an SU(2) sub-group of a larger gauge symmetry for the vector reggeons, a new phenomenon emerges that has far-reaching ramifications.

In addition to the cancellation of reggeization divergences, an overall infra-red scaling divergence appears in SU(2) color zero amplitudes that contain combinations of ``anomalous wee gluon reggeons'' coupled via {\bf anomaly pole reggeon vertices}. The wee gluon reggeon combinations are multi-reggeon (non-local) generalizations of the well-known anomaly current. They are massless, with overall SU(2) color zero, anomalous color parity ($\neq$ signature), and all carrying zero transverse momentum. The divergence selects physical reggeon amplitudes. Because it is an overall divergence, the states and interactions appear simultaneously and so vacuum-like ``universal wee partons'' are necessarily present in all states and interactions.
 
Anomaly poles\cite{arw4} are produced by chirality transitions in massless fermion triangle diagrams. They appear in triangle diagram reggeon vertices  
that contribute in non-planar multi-regge limits where bound-state amplitudes appear.
The vector and axial vector couplings involved have two sources. Firstly, when multiple massless vector (gluon) reggeons are present in each of three distinct (non-planar) rapidity channels and couple into the same fermion triangle vertex, orthogonal $\gamma$-matrices accumulate to produce a $\gamma_5$. Secondly, on-shell {\bf massive} vector exchange between a massless fermion/antifermion reggeon pair produces an effective vector vertex that couples to triangle anomalies. A pseudoscalar anomaly pole can couple, therefore, to a reggeon state composed of a fermion/antifermion pair and anomalous wee gluons.

As the full gauge symmetry is restored (in the specific cases we discuss below) there is an exploitation of the Gribov ambiguity. The zero momentum transverse component of the on-shell massive vector exchanged between the fermion/antifermion pair, in pseudoscalar anomaly poles, does not decouple as it becomes massless. These anomaly poles become the pseudoscalar Goldstone bosons associated with chiral symmetry breaking. They are also essential components of the fermion bound-states.
Therefore, the Gribov ambiguity is fundamental in the formation of the spectrum of physical states that appear in the unbroken gauge theory.
 
\subsection{Full Gauge Symmetry With Only Massless Fermions}

The transverse momentum cut-off has to be removed first if the full massless limit is to be well-defined. This can be done smoothly if the scalar sector to be decoupled is asymptotically free. To achieve this, many fermions are needed to slow down the evolution of the gauge coupling - the simplest example being QCD with the asymptotic freedom constraint saturated. An even more restrictive constraint is that the anomaly pole coupled infra-red divergence has to be preserved in  all orders of perturbation theory. This requires that the fermions are massless and produce an infra-red fixed point. 

\subsubsection{Massless QCD$_S$ $\equiv$ 6 triplet quarks + 2 sextet quarks }

This\cite{arw2,arw3,arw4} is the only, physically relevant, possibility to saturate QCD with massless fermions. A pair of color sextets (that can be associated with electroweak symmetry breaking) is added to the physical triplet sector. We initially consider color superconducting QCD$_S$ - with the gauge symmetry broken to SU(2) by an (asymptotically free) color triplet scalar field.
As implied above, when the overall divergence is subtracted to define physical amplitudes we obtain an SU(2) color confining ``parton model''
in which anomalous wee gluons  provide a vacuum-like ``wee parton condensate''
within states and interactions. Anomaly poles produce chiral ``pion'' 
poles (containing both fermion pairs and fermion/antifermion pairs) and also couple wee gluons present in distinct reggeon channels. There is both confinement and chiral symmetry breaking! 

The exchanged pomeron interaction is a massive, SU(2) singlet, gluon reggeon accompanied by the wee gluons. It is exchange degenerate with the gluon reggeon and can be identified with the supercritical pomeron. Consequently, the massless limit restoring SU(3) color is described\cite{arw2,arw3,arw4} by the Critical Pomeron. Many experimentally desirable features appear.
\begin{itemize} 
\item{\it Bound-states are pseudoscalar mesons and baryons containing triplet and sextet quarks, but with no hybrid sextet/triplet states,} 
\item{\it Anomaly color factors give larger masses and x-sections for sextet states.}
\item{\it There are no glueballs, no BFKL pomeron, and no odderon.}
\item{\it The Critical Pomeron gives maximal parton model factorization.}
\end{itemize}

The infra-red fixed-point implies off-shell amplitudes have to be scale invariant and so can not contain massive states. Therefore, if a bound-state S-Matrix containing massive particles exists, there must be {\bf no off-shell correlation functions!} Indeed, the existence of off-shell amplitudes seems very unlikely to be 
consistent with the anomaly pole dynamics. Nevertheless, a huge chiral symmetry implies the S-Matrix necessarily has many massless Goldstones that would pose a very serious, if not insoluble, infra-red problem. 

At this point, it seems questionable that the Critical Pomeron can appear in a physical S-Matrix that has only massive states (as in the real world!) Fortunately, the electroweak interaction provides a way out of this dilemma. 
A remarkable result emerges when we consider combining the electroweak interaction with QCD$_S$. We discover a unique theory\cite{arw3,arw4,kw,arw5,arw6,arw7,arw8} that is again massless, asymptotically free, and saturated with fermions, but has left-handed couplings to all fermions.

\subsubsection{QUD}

Asking for asymptotic freedom and no short-distance anomaly, massless QCD$_S$ and the electroweak interaction together embed uniquely in \begin{center}
{\bf QUD $\equiv$ SU(5) gauge theory with left-handed massless fermions in the 
$5 \oplus 15 \oplus40 \oplus 45^*$ representation.}\end{center} 
\noindent The asymptotic freedom constraint is again saturated  
and there is an infra-red fixed-point. Now, however, there are no exact chiral symmetries and so all bound-states can acquire S-Matrix masses via interactions. Amazingly, 
\begin{center}{\bf QUD has the additional structure needed to generate a bound-state S-Matrix that produces all the physics of the Standard Model via massless fermion anomaly dynamics.}
\end{center}
The only ``beyond the Standard Model" elements are a dark matter sector and neutrino masses, both of which are extremely welcome.  

There are three ``generations'' of both elementary leptons and elementary triplet quarks, and the theory is vector-like with respect to SU(3)xU(1)$_{em}$. The SU(2)xU(1) quantum numbers are not quite right but, in the S-Matrix constructed\cite{arw1,arw2,arw6} via multi-regge theory, all elementary fermions are confined and only Standard Model interactions and states emerge. The color sextet sector provides ``sextet pions'' that produce electroweak symmetry breaking and sextet baryons - with the sextet neutron and antineutron providing stable, massive, {\bf dark matter} particles. There is also a color octet sector that is responsible, via large $k_{\perp}$ anomalies, for the generation structure of the physical states. We can briefly summarize how we obtain QUD bound-state amplitudes from reggeon diagrams as follows.

\subsubsection{QUD Reggeon Diagrams}

QUD reggeon diagrams are initially defined with massive reggeons.
Fermion masses are removed first, then the gauge boson global symmetries are restored 
through a sequence of fundamental representation scalar decouplings   
 \newline $~$
 \newline
 \centerline{ $\rightarrow SU(2)_C~, \rightarrow  SU(4),~
  \lambda_{\perp} \to \infty,~ \rightarrow SU(5)$}
 
 \noindent The last scalar is asymptotically free and so the $\lambda_{\perp} \to \infty$ limit can be taken between the SU(4) and SU(5) limits. The SU(2)$_C$ symmetry  restoration produces anomaly poles, but with a difference. As in QCD$_S$,  the chirality transitions left by the fermion mass removal do not conflict with the {\bf vector} gauge symmetry but, crucially, they do break the {\bf non-vector} part of the gauge symmetry. The SU(5) symmetry is broken to SU(3)$_C\otimes$U(1)$_{em}~$, but {\bf in the reggeon anomaly vertices only.} The final ``universal wee partons'' are combinations of vector coupling anomalous wee gauge bosons in the adjoint SU(5) representation. The global SU(5) symmetry is restored within the reggeon interactions between anomaly vertices. The {\bf Gribov ambiguity is essential} for the breaking, {\bf via the anomaly vertices,} of the SU(5) symmetry in the bound-state S-Matrix and, in particular, for the electroweak vector boson masses that are generated by mixing with the sextet pions.

 \subsubsection{QUD Interactions and States}
 
 As I have described in my papers, much of the Standard Model is clearly present in the QUD bound-state amplitudes
 extracted from the reggeon diagrams. Indeed, most likely, it is all there but much remains to be better understood. Standard Model vector interactions between bound-states are - 
 \begin{itemize}
 \item{\it Critical Pomeron $\approx$ SU(3) gluon reggeon + wee gauge bosons.}
 \item{\it Photon $\approx$ a U(1)$_{em}$
 gauge boson + wee gauge bosons} 
 \item{\it Weak Interaction $\approx$ l-handed gauge bosons mixed with  sextet pions \newline + wee gauge bosons.}
\end{itemize}
The elementary QUD coupling stays very small because of the infra-red fixed-point.
The Standard Model couplings are enhanced by anomaly color factors that imply
$$~\alpha_{\scriptscriptstyle QCD} ~> ~\alpha_{\scriptscriptstyle em}~>>~ \alpha_{\scriptscriptstyle QUD} \sim
 \frac{1}{120}$$
 
 Bound-states contain elementary fermions that combine with color octet ``pions'' (present at infinite light-cone momentum as an anomaly contribution) to form SU(5) singlets. Standard Model generations of physical hadrons and leptons appear. Anomaly vertex mixing, combined with fermion and wee parton color factors, produces a wide range of mass scale, with the very small QUD coupling, presumably, the origin of small neutrino masses. In general, anomaly color factors imply 
 $$M_{\scriptstyle hadrons} >> M_{\scriptstyle leptons} >>  M_{\scriptstyle \nu 's} ~\sim~
  \alpha_{\scriptstyle QUD}$$
  
 There are three generations of lepton bound states. Each physical lepton contains three elementary leptons, with two of them originating from an anomaly pole. The electron is almost elementary since, in effect, the anomaly pole disturbs the Dirac sea minimally. The muon has the same constituents, but in a different dynamical configuration that will obviously generate a significant mass. 
 
 We can summarize hadronic sector results, very briefly, as follows.  
  \begin{itemize}
 \item{\it Two triplet generations of Standard Model hadrons that mix.}
 \item{\it The physical b quark is a mixture of all three QUD generations}.
 \item{\it Sextet pion/vector boson mixing $\to$ electroweak symmetry breaking.} 
 \item{\it Sextet neutrons are stable and provide dark matter.}
 \item{\it l-handed ``top quark'' mixes with  exotic quarks, no low mass states.}
 \item{\it Triplet/sextet mixing gives mixed-parity $\eta_3$ and $\eta_6$ scalars.} 
 \item{\it $\eta_6$ gives Standard Model ``top physics'' with sextet mass scale.}
 \item{\it $\eta_3$ could  be  the ``Higgs boson''.}
 \end{itemize}
 More details of these results can be found in my papers. 
 Theoretical virtues for the QUD origin of the Standard Model are also listed in several of the papers. A brief list is
 \begin{itemize}
 \item{\it Strong/electromagnetic/weak interaction parities explained.}
 \item{\it Confinement/chiral~symmetry~breaking/the parton model/the Critical Pomeron all appear in QCD.}
 \item{\it The massless photon partners the ``massless'' Critical $~\pom~$.}
 \item{\it Anomaly vertices/wee parton color factors give a wide range of 
 scales and masses, with neutrino masses $\sim$ very small $\alpha_{QUD}$.}
 \item{\it The only new physics is a high mass strong interaction giving electroweak symmetry breaking and dark matter.}
 \item{\it Particles and fields are truly distinct. Hadrons and leptons have equal status.}
 \item{\it Symmetries and masses are dynamical S-Matrix properties. There are no  off-shell amplitudes and there is no Higgs field.}
 \item{\it Einstein gravity is induced with zero cosmological constant. Gravity is not quantized !! Particle Physics is described by an S-Matrix. }
\end{itemize}
     
 Obviously, it would be incredible if the Standard Model has the underlying simplicity of QUD. It is important to emphasize again that there is no freedom for variation. It is an ``all or nothing'' explanation of the origin of the Standard Model which predicts the, apparently observed, ``nightmare scenario'' of a ``Higgs boson'' produced at the LHC without other new short-distance physics! As is also described in my papers, there is much suggestive cosmic ray and  accelerator experimental evidence for the electroweak scale strong (not short-distance) QCD interaction of the sextet quark sector. Particularly striking are\cite{arw9} the latest AMS-02 results
 and the most recent observation of dark matter strong self-interaction properties.
 
\section{Conclusions and Outlook}

I was led to the QUD bound-state S-Matrix by looking for a gauge theory origin for the Critical Pomeron - an S-Matrix phenomenon formulated far from the framework of quantum field theory. It could well be, as I have suggested\cite{arw1}, that the Gribov copies are a major reason why the path-integral fails to provide a non-perturbative formulation of a non-abelian gauge theory. In this case, the field theory can be well-defined perturbatively and yet the particle S-Matrix could be the only well-defined non-perturbative element that exists. 

I have argued that the Critical Pomeron high-energy S-Matrix is uniquely associated with QUD. I have emphasized that QUD is self-contained and must reproduce the established physics of the Standard Model, or else it is simply wrong ! The potential scientific and aesthetic importance of QUD,
as an underlying unifying massless field theory, is overwhelming. So can the necessity for the underlying theory to be QUD be seen without the Critical Pomeron? 

The infra-red fixed-point is required to 
enhance infra-red fermion anomaly interactions and for the color-superconductivity starting point that provides the exploitation of the Gribov ambiguity. The vector gauge group has to be as large as SU(3) to produce a universal wee gluon distribution, as well as an infinite momentum ``parton model'' giving an ultra-violet finite S-Matrix. 
If the vector gauge group is larger than SU(3), 
the universality of the wee gluon distribution is lost. Left-handed interactions, that acquire a mass via infra-red anomalies, and also generate bound state masses, can be 
added. Asking that all bound-states acquire masses, plus no short-distance
anomaly leads, perhaps uniquely, to QUD. So, it could be that
\begin{center}
{\it The Standard Model is reproducing the unique, unitary, S-Matrix~??}
\end{center}

If it is eventually eestablished that there is indeed a unique S-Matrix and the Gribov discoveries of reggeon unitarity and gauge field copies are the fundamental elements that determine it's existence then it will be a dramatic addition to the legacy of Vlodya Gribov\footnote{Even though it is very doubtful that he would have actually endorsed my argumentation.}. 

\section{Epilogue}

As is surely evident, there is an enormous amount of work needed to fully develop my multi-regge construction of the QUD S-matrix into a practical formalism. Far more than I could ever do on my own\footnote{Despairing over the situation I wrote to Bj, {\it ``What I really need and, of course, have zero chance of getting, is a ``young Bjorken'' to work with me!} Bj replied {\it ``Much better would be a young Gribov''} I replied {\it ``If only !!!!!!''}}
. I have given\cite{arw8} a partial list of the multi-regge developments that are needed. Once the diagrams have been fully characterized, there is still the introduction of scales, including the regularization of the anomaly vertices and the description of the Critical Pomeron limit. Given the disappearance of the intense scientific environment that produced Gribov (a bizarre consequence, mainly, of Soviet repression), and the current focus of the field on inventive intellectual forays that are far away from unsolved hard core physics problems, it is difficult to imagine where and when the effort will be made to develop all the details of the physical picture I have outlined. 

The most serious obstacle is, of course, that a major change of theory paradigm is involved. I am saying that quantum fields exist only as short-distance, gauge-dependent, fluctuations for which the only manifest physical reality is the particle S-Matrix. This is in stark contrast with the concepts and philosophy of the vast majority of today's "Beyond the Standard Model" research. Nevertheless, if it is reality, it will eventually emerge and, I at least, can imagine that with his scientific background, Gribov would not have been entirely unsympathetic!

\end{document}